\def \vc #1{{\mbox{\boldmath $#1$}}}

\def\thetag{{\vc \theta}}

\def\alphag{{\vc \alpha}}
\def\deltag{{\vc \delta}}
\def\gammag{{\vc \gamma}}

\documentclass[]{spie}  
\usepackage[dvips]{graphicx}

\title{Wide-field cosmic shear surveys} 

\author{Mellier Y.\supit{a,b,c}, van Waerbeke L.\supit{a},
  Bertin E.\supit{a,b,c}, Tereno I.\supit{a} and 
 F. Bernardeau\supit{c} 
\skiplinehalf
\supit{a}IAP, 98 bis Blvd Arago, 75014 Paris, France \\
\supit{b}Obs. de Paris/LERMA, 77 Av. Denfert-Rochereau, 75014 Paris, France \\
\supit{c}TERAPIX-IAP, 98 bis Blvd Arago, 75014 Paris, France \\
\supit{d}SPhT, CE Saclay, 91191 Gif sur Yvette cedex, France.
}

\authorinfo{Further author information: (Send correspondence to Y. Mellier)\\Y. Mellier: E-mail: mellier@iap.fr, Telephone: (+33) 1 44 32 81 40\\  
}

\begin{document} 
  \maketitle 

\begin{abstract}
We present the current status of cosmic shear  based on 
   all surveys done so far. Taken together, they cover more about 70 deg$^2$ 
  and concern more than 3 million galaxies with accurate shape measurement.
  Theoretical expectations, observational results and their cosmological 
 interpretations are discussed
  in the framework of standard cosmology and CDM scenarios.
   The potentials of the
next generation cosmic shear surveys are discussed.

\end{abstract}


\keywords{Cosmology, Large Scale Structure, Gravitational Lensing, Weak Lensing, Surveys}

\section{INTRODUCTION}
\label{sect:intro}  
Gravitational lensing produces  
 distortion of light beams which modifies the image shapes of
 background galaxies.
  In a Friedman-Robertson-Walker metric,
   and for stationary, weak gravitational
fields, the deflection angle  writes
\begin{equation}
  \hat\alphag
  = \frac{2}{c^2}\,\int\nabla_\perp\Phi\, \ {\rm d}l \ ,
\label{deflec}
\end{equation}
where $c$ is the celerity and $\Phi$ the
  3-dimension gravitational potential.
   In general, the
deflection angle is small and lenses can be approximated as thin
   gravitational systems. This simplifies the relation between
  the source ($S$) and image ($I$) positions according to the simple
  geometrical ``lens equation'':
\begin{equation}
\thetag^I=\thetag^S+{D_{LS} \over D_{OS}} \hat\alphag(\thetag^I) \ ,
\label{lensequa}
\end{equation}
where $D_{ij}$ are angular diameter distances.\\
Equations (\ref{deflec}) and (\ref{lensequa}) express how
  lens properties depend on dark matter distribution and
  on cosmological models.
  From an observational point of view, gravitational lenses
  manifest as image multiplication of galaxies or quasars, strong and weak
    distortions of galaxy shape or
  transient micro-lensing effects.  These effects, as well as 
   the  time delays attached to image multiplications 
 are exploited
 in order to  probe the geometry and matter/energy content 
  of the Universe or to observe high-redshift 
galaxies
 (see \citenum{blandnar92}, \citenum{fortmel94}, \citenum{mel99}, 
  \citenum{bs01} for reviews)
 \ \ .
\\
In the
case of weak gravitational lensing, useful approximations
   relating observed shapes of galaxies to gravitational
  shear can be made.
 In general, the image magnification is  characterized by the 
   convergence, $\kappa$,  
   and by the shear components $\gamma_1$ and $\gamma_2$:
\begin{equation}
 \kappa = \frac{1}{2}\,(\varphi_{,11}+\varphi_{,22})  ; \ \  \ \ \ \ \
  \gamma_1(\theta) =
  \frac{1}{2}\left(\varphi_{,11}-\varphi_{,22}\right) ; \ \ \ \ \ \ \
  \gamma_2(\theta) = \varphi_{,12} = \varphi_{,21}
\label{convshear}
\end{equation}
 {\rm where } the $\varphi_{,ij}$ are the second derivatives 
  of $\varphi$ with respect to the $i,j$ coordinates and
\begin{equation}
\varphi(\theta) = \frac{2}{c^2}\ \frac{D_{LS}}{D_{OS}D_{OL}}\,
  \,\int\,\Phi(D_{OL}\theta,z)\, \ {\rm d}z\;.
\label{projpoten}
\end{equation}
The shear applied to lensed galaxies
  increases  their ellipticity along a direction perpendicular to
  the gradient of the projected potential.  The lens-induced distortion
   $\deltag$ can then be 
    evaluated from the  shape of galaxies as it can be observed from
  the components of their  surface brightness second moment 
   $M_{ij}= \displaystyle{
 {\int
I\left(\thetag\right) \theta_i \ \theta_j
 \ {\rm d}^2\theta \over \int I(\thetag) \ {\rm d}^2\theta }} \ :
$
\begin{equation}
\deltag ={2 \gammag \ (1-\kappa)
\over
(1-\kappa)^2+\vert \gamma \vert^2} \ =\left(\delta_1={M_{11}-M_{22}
\over Tr(M)} \ ;  \ \delta_2={ 2 M_{12} \over Tr(
M)} \right) \ .
\label{distor}
\end{equation}
In the weak lensing regime, 
  the relation between the distortion and the gravitational shear  
 simplifies to $\deltag \approx 2 \gammag $, so
 in principle  ellipticity of galaxies, as measured
from the  second moment components, provides a
 direct estimate of the gravitational shear.
   The
  ``shear map'' can then be used to reconstruct the ``mass map''
 at any galaxy position.
     However,  since each galaxy has
its  own intrinsic ellipticity, and since the background galaxies
   only sparsely sample the sky, the final shear-induced
   ellipticity map is contaminated by 
  shot noise  and cannot have infinite resolution.
  \\
The use of ellipticities of galaxies for weak lensing statistics
 has three important applications: galaxy-galaxy lensing,
  mass reconstruction of clusters of galaxies
  and  gravitational distortion
  induced by large-scale structures of the universe. Each
  permits to recover properties of the dark matter located in gravitational
   systems as well as the mass density of the universe.
 In the following we only focus on  
  weak lensing analysis applied to large-scale structures, namely the cosmic 
 shear.
\section{Motivations for Cosmic Shear surveys}
Cosmic shear refers to the  weak distortion  of light bundles 
  produced  by the cumulative effects of mass inhomogeneities in the 
universe.  Although theoretical cosmological weak lensing studies 
started more than 30 years ago, it is only recently that this
 topic  moved to an observational cosmology tool.
  Progresses in the field are going incredibly fast, despite
  observational and technical difficulties, and put cosmic shear
  as CMB was only few years ago.
  These remarkable progresses were  possible because  wide
field surveys with panoramic CCD cameras grow 
  rapidly and can be handled with present-day computing facilities.
\\
Light propagation through an inhomogeneous universe accumulates
 weak lensing effects over Gigaparsec distances.
 Assuming structures formed from gravitational growth of
    Gaussian fluctuations, the shape and amplitude 
  of cosmological weak lensing as function of angular scale can be predicted
from Perturbation
  Theory.  To first order, the convergence $\kappa(\thetag)$ at
angular
position $\thetag$ is given by the line-of-sight integral
\begin{equation}
\kappa(\thetag)={3 \over 2} \Omega_0  \int_0^{z_s} n(z_s)  {\rm d}z_s
\int_0^{\chi(s)}  {D\left(z,z_s\right) D\left(z\right) \over
D\left(z_s\right)}  \delta\left(\chi,\thetag\right) \
\left[1+z\left(\chi\right)\right]  {\rm d}\chi
\end{equation}
where $\chi(z)$ is the radial distance out to redshift $z$, $D$ the
angular diameter distances, $n(z_s)$ is the redshift
distribution of the sources.
 \ \ \   $\delta$ is
the mass density contrast responsible for the deflection at redshift
 $z$. Its amplitude at a given redshift  depends on the properties  of the
 (dark) matter power spectrum and its evolution with look-back-time.
\\
The cumulative weak lensing effects of
structures induce a shear field
  which is primarily related to the   power spectrum of the projected
mass density, $P_\kappa$.
   Its statistical properties can be recovered by
    the
shear top-hat variance (\citenum{me91,b91,k92}),
\begin{equation}
\langle\gamma^2\rangle={2\over \pi\theta_c^2} \int_0^\infty~{{\rm
d}k\over k} P_
\kappa(k)
[J_1(k\theta_c)]^2,
\label{theovariance}
\end{equation}
the aperture mass variance (\citenum{k94,sch98})
\begin{equation}
\langle M_{\rm ap}^2\rangle={288\over \pi\theta_c^4} \int_0^\infty~{{\rm
d}k\over k^3}
P_\kappa(k) [J_4(k\theta_c)]^2,
\label{theomap}
\end{equation}
 and the shear correlation
function (\citenum{me91,b91,k92}):
\begin{equation}
\langle\gamma\gamma\rangle_\theta={1\over 2\pi} \int_0^\infty~{\rm d} k~
k P_\kappa(k) J_0(k\theta),
\label{theogg}
\end{equation}
where $J_n$ is the Bessel function of the first kind.  $P_\kappa(k)$ is 
 directly related to the 3-dimension power spectrum of the dark matter 
   along the line of sight, $P_{3D}$:
\begin{equation}
P_\kappa(k)= {9 \over 4} \Omega_m^2 \ \int_0^\infty \ P_{3D}\left(
{k \over D_L\left(z\right)}, z\right) \ W(z,z_s) \ {\rm d}z \ ,
\label{p3d}
\end{equation}
where $ W(z,z_s)$ is an efficiency function which depends on the redshift 
  distribution of sources and lenses.
 Therefore,  in principle an inversion permit to reconstruct the
  3-dimension power spectrum of the dark matter from the 
  weak distortion field. \\
 Higher order statistics, like  the skewness
of the convergence, $s_3(\theta)$, can also be computed.
  They probe non Gaussian
features in the projected mass density field, like
    massive clusters or compact groups of galaxies (\citenum{bern97}$^,$
\citenum{jain97}).  The amplitude
 of cosmic shear signal and its sensitivity to cosmology
  can be  illustrated in the fiducial case of a power
law mass power spectrum with no cosmological constant and
a background population at a single redshift
$z$. In that case  $<\kappa(\theta)^2>$ and $s_3(\theta)$ write:
\begin{equation}
\label{eqvar}
<\kappa(\theta)^2>^{1/2} =<\gamma(\theta)^2>^{1/2} \approx 1\% \
 \sigma_8 \ \Omega_m^{0.75} \
z_s^{0.8
} \left({\theta \over  1'}\right)^{-\left({n+2 \over 2}\right)}  \ ,
 \
\end{equation}
and
\begin{equation}
\label{eqs3}
s_3(\theta)={\langle\kappa^3\rangle\over \langle\kappa^2\rangle^2}
\approx 40 \
  \Omega_m^{-0.8} \ z_s^{-1.35}  \ ,
\end{equation}
 where $n$ is the spectral index of
the power
spectrum of density fluctuations. Therefore, in principle the
      degeneracy between $\Omega_m$ and $\sigma_8$ can be  broken
   when both the variance and the skewness of the convergence
  are measured.
\section{Detection of weak distortion signal}
\subsection{An observational challenge}
\begin{figure}[t]
\begin{center}
\includegraphics[width=11cm,angle=270]{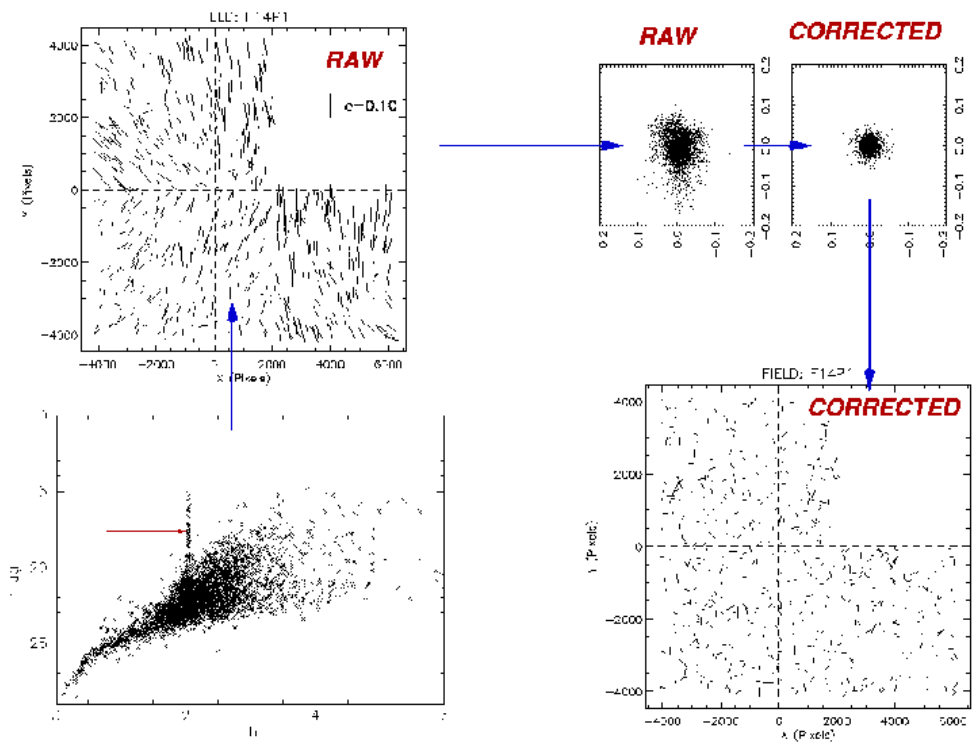}
\end{center}
\caption[]{Basic scheme for the non-lensing distortion correction of 
 cosmic shear surveys. Stars of each field are first selected in the 
vertical branch of the size-magitude plots. The  PSF (Raw)
  is then mapped over the whole field which permits to remove the 
   non-lensing distortion anywhere in the field from simple interpolations. 
  The corrected stars provide an estimated of the residual (Corrected).
}
\label{psfcorrection}
\end{figure}

Eq. (\ref{eqvar}) shows that  the amplitude of weak lensing signal is
of the order of few percents, which is
   much smaller than the
  intrinsic dispersion of ellipticity distribution of galaxies.  Adopting the 
  null hypothesis that no gravitational weak lensing signal is present, 
  we can predict the expected limiting shear  
  if only shot noise and sampling are taken into account (in particular,
   we assume there is no systematic residual errors associated to 
  image processing or PSF corrections as those discussed later). 
  For a $3\sigma$ shear variance limiting detection it writes :
\begin{equation}
\label{survey}
<\gamma(\theta)^2>_{limit}^{1/2} = 1.2\%  \ \left[{A_T \over 1
\ deg^2}\right]^{-{1 \over 4}} \times \left[{\sigma_{\epsilon_{gal}}
\over 0.4} \right] \times
 \left[{n \over 20 \ gal/arcmin^2}\right]^{-{1 \over 2}} \times
 \left[{\theta \over 10'}\right]^{{-{1 \over 2}}} \ ,
\end{equation}
where $A_T$ is the total area covered by the survey, $\sigma_{\epsilon_{gal}}$ 
 is the intrinsic ellipticity dispersion of galaxy and $n$ the galaxy number 
  density of the survey.
 By comparing with  Eq.(\ref{eqvar}) one concludes 
  that a cosmic shear survey covering $1$ deg$^2$ up to 
  a reasonable depth corresponding to 20 gal./arcmin$^2$ should in principle 
  be the minimum requirements
   to measure a cosmic shear signal. However, when gravitational shear 
  is included these limits change as function of  
   the angular scale and of cosmological parameters. 
  Our simplistic estimate  turns out to be too pessimistic on
  small scales because non-linear growth of perturbation producing 
  clusters, groups and galaxies amplify the cosmic shear signal by a factor 
  of 2 to 10, depending on cosmologies (\citenum{bern97}) \ .
 Van Waerbeke et al (\citenum{vwal99}) explored which  strategy would be
best suited to probe statistical properties  of such a small signal. 
  Their results are illustrated on 
   Table \ref{tab1} and show  that the   variance of $\kappa$
 can already provide a valuable cosmological 
  information, even with a survey covering about 1 deg$^2$. However, 
   for the skewness one needs at least 10 deg$^2$. Furthermore,
   more than 100 deg$^2$ must be observed in order to
   uncover  information on $\Omega_{\Lambda}$
  or  the shape
  of the power spectrum over scales larger than 1 degree.
\begin{table}[t]
\caption{\label{sizesurvey}Expected signal-to-noise ratio on the
the variance and the skewness of the convergence for two
  cosmological models.  In the first
column, the size of the field of view (FOV) is given. The
signal-to-noise
 ratio is computed from the simulations done by
van Waerbeke et al  (1999). }
\begin{center}
\begin{tabular}{|l|c|c||c|c|}
\hline
\multicolumn{5}{|c|}{$z_s=1$, Top Hat
filter , $n=30$ gal.arcmin$^{-2}$} \\ \hline
\hline
  FOV &
\multicolumn{2}{|c|}{ {S/N Variance}} &
\multicolumn{2}{|c|}{ {S/N Skewness}} \\
\cline {2-5}
 (deg.$\times$deg.) &  $\Omega_m=1$  &
 {$\Omega_m=0.3$}  &
 {$\Omega_m=1$}  &
 {$\Omega_m=0.3$}  \\ \hline
 1.3 \ $\times$ \ 1.3  &  7  &  5 &  1.7 &   2 \\ \hline
 2.5 \ $\times$\ \ 2.5  & 11  & 10 & 2.9 &  4 \\ \hline
 5.0 \ $\times$ \ 5.0  & 20  & 20 & 5 &  8\\ \hline
 10.0 \ $\times$ \ 10.0  & 35 & 42 & 8 &  17\\ \hline
\end{tabular}
\label{tab1}
\end{center}
\end{table}
\\
The shape of objects may also  be alterated by atmospheric turbulence, 
 optical and atmospheric distortions as well as unexpected telescope 
  oscillations or CCD charge   transfer efficiency.  Non-cosmological 
 signals reach amplitudes  one order of magnitude higher than cosmic 
shear and removing them is indeed the most challenging issues of these 
surveys. Fortunately, these alterations also modify the shape of stars, 
  hence lensing and non-lensing contributions to galaxy shapes can be 
  separated.   Over the past decades many works focussed on the corrections
  of non-lensing contributions. All follow the basic scheme
  shown in Figure\ref{psfcorrection}, but use different  star 
   shape analyses and correction techniques (see \citenum{mel99} and 
  \citenum{bs01} for comprehensive reviews). 
 The reliability of artificial anisotropy corrections
 has been discussed and tested at 
length (\citenum{erben01}, \citenum{vwal00}, \citenum{bacon00}, \citenum{bacon01})\ .  In particular, Erben et al (\citenum{erben01}) have 
 demonstrated 
  that  the so-called KSB method (\citenum{ksb})
   is able to correct PSF anisotropy very well 
  and allows astronomers to detected cosmic shear amplitude down to 1\% 
  level with a 10\% relative accuracy, even if the non-lensing contribution 
  is as large as 10\%.

\subsection{Detection}
\begin{figure}[t]
\begin{center}
\includegraphics[width=10cm,angle=270]{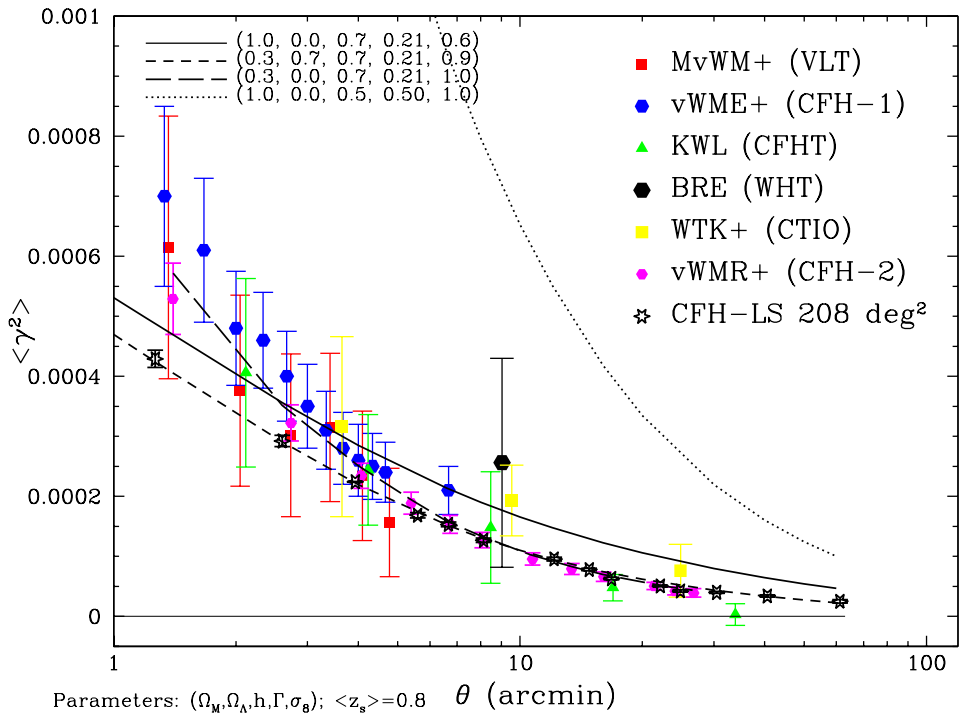}
\end{center}
\caption[]{Top hat variance of shear as function of angular scale from
  6 cosmic shear surveys. The acronyms of the right refer to the 
  references reported in Table 2. The  open black stars
 illustrate the expected performances of the CFHT-LS
  which will start by 2003 with Megacam at CFHT. This is 
 the expected signal from  the ``Wide Survey'' which 
  will cover about  200 deg$^2$ up to $I_{AB}=24.5$. As one can see, for 
most points the errors are smaller than the stars.
}
\label{sheartop}
\end{figure}

As we discussed in the previous section,
 on scale significantly smaller than one degree,
non-linear structures dominate and increase the amplitude of the lensing
signal, making its measurement easier.
  Few teams  started such surveys
  during the past years and succeeded to get a significant signal.
 Table \ref{tabcs} lists some published results.
  Since each group used different telescopes,
  adopted different observing strategy and
 used different data analysis techniques,  one can figure out
  their reliability by comparing these surveys.
\begin{table}
\begin{center}
{\small
\caption{Present status of cosmic shear surveys with published
results.
}
\label{tabcs}
\begin{tabular}{lcccl}\hline
Telescope& Pointings & Total Area & Lim. Mag. & Ref.. \\
\hline
CFHT & 5 $\times$ 30' $\times$30'& 1.7 deg$^2$ & I=24. &
\citenum{vwal00}[vWME+]\\
CTIO & 3 $\times$ 40' $\times$40'& 1.5 deg$^2$ & R=26. &
\citenum{wit00a}[WTK+]\\
WHT & 14 $\times$ 8' $\times$15'& 0.5 deg$^2$ & R=24. &
\citenum{bacon00}[BRE]\\
CFHT & 6 $\times$ 30' $\times$30'& 1.0 deg$^2$ & I=24. &
\citenum{kais00}[KWL]\\
VLT/UT1 & 50 $\times$ 7' $\times$7'& 0.6 deg$^2$ & I=24. &
\citenum{mao01}[MvWM+]\\
HST/WFPC2 & 1 $\times$ 4' $\times$42'& 0.05 deg$^2$ & I=27. &
\citenum{rhodes01}\\
CFHT & 4 $\times$ 120' $\times$120'& 6.5 deg$^2$ & I=24.
&\citenum{vwal01}[vWMR+]\\
HST/STIS & 121 $\times$ 1' $\times$1'& 0.05 deg$^2$ & V$\approx 26$
& \citenum{hammerle01}\\
CFHT & 5 $\times$ 126' $\times$140'& 24. deg$^2$ & R=23.5& \citenum{hoek01a}
\\
CFHT & 10 $\times$ 126' $\times$140'& 53. deg$^2$ & R=23.5& \citenum{hoekstra02}
\\
CFHT & 4 $\times$ 120' $\times$120'& 8.5 deg$^2$ & I=24.&
\citenum{pen01}\\
HST/WFPC2 & 271 $\times$ 2.1 $\times$ 2.1 & 0.36 deg$^2$ & I=23.5 & \citenum{ref02}
 \\
Keck+WHT & 173 $\times$ 2' $\times$ 8' & 1.6 deg$^2$ & R=25 & \citenum{bacon02} \\
 & +13 $\times$ 16' $\times$ 8' &  &  &  \\
 & 7 $\times$ 16' $\times$ 16' & &  &  \\
\hline
\end{tabular}
}
\end{center}
\end{table}
Figure \ref{sheartop} show that they are all
   in very good
agreement\footnote{ The Hoekstra et al (\citenum{hoek01a}) data are missing
   because depth is
  different so the
sources are at lower redshift and the amplitude of the shear
  is not directly comparable to other data plotted}.  This  is
  a convincing demonstration that the
  expected correlation of ellipticities is real.
\subsection{Cosmological nature of the signal}
The cosmological origin of the coherent distortion signal detected 
   by all these surveys is not obvious. 
 Even if a cosmological signal is present in the data, it could be
 strongly contaminated by unknown  systematic contributions
   or residuals  systematic  errors 
   produced by the PSF corrections discussed above. 
 An elegant way
   to check whether corrections are correctly done and to
   confirm the   gravitational nature of the signal is
  to decompose the signal into  E- and B- modes.
   The E-mode is a gradient term which 
 contains signal produced by gravity-induced distortion. On the other hand, 
the B-mode is a pure curl-component, so it
  only contains intrinsic ellipticity correlation or
systematics residuals. Both modes have been extracted using the
aperture mass statistics by van Waerbeke et al (\citenum{vwal01},
 \citenum{vw02}) 
 \ and
 Pen et al (\citenum{pen01}) in the VIRMOS-DESCART\footnote{
For the VIRMOS spectroscopic survey, see (\citenum{olf02}). 
  For the VIRMOS-DESCART survey see http://terapix.iap.fr/Descart/}\footnote{Data of the VIRMOS-DESCART survey were processed and analysed at the
 TERAPIX data center: http://terapix.iap.fr}
     survey as well as
  by Hoekstra et al (\citenum{hoekstra02}) in the Red Cluster Sequence survey.
 In both samples, the E-mode dominates the signal, although a small
  residual is detected in the B-mode on small scales for the 
  RCS sample and on large scales for the VIRMOS-DESCART sample. These 
  residuals still need further investigations. This strongly supports the
   gravitational origin of the distortion.
\\
An alternative to a gravitational lensing  origin to the signal could be an
   intrinsic correlations of ellipticities of galaxies produced
  by proximity effects (tidal torques).   Several
  recent  numerical and theoretical  studies 
 (\citenum{crit00}, \citenum{mackey01})
 have addressed this point. 
   From these first investigations, it seems 
    that intrinsic correlations are negligible on scales beyond
  one arc-minute, provided the survey is deep enough, 
  but this critical point is still debated. Nevertheless, 
   we hope that deep surveys are reliable since in that case,
  most lensed galaxies along a line of sight are spread over
 Gigaparsec scales and have no physical relation with its apparent
  neighbors.
  Hence, since most cosmic survey are deep, we expect they are almost
    free of intrinsic correlations.
\section{Cosmological interpretations}
\subsection{2-point statistics and variance}
A comparison of the top-hat shear variance 
   with some realistic cosmological models
 is ploted in Figure \ref{sheartop}. Its amplitude 
  has been scaled using photo-$z$ which gives $<z> \approx 1$.
 On this plot, we see that
 standard COBE-normalized CDM is ruled  at a 10$-\sigma$  confidence
level.  However,
  the degeneracy between $\Omega_m$ and
   $\sigma_8$  discussed in the previous section still hampers
  a strong discrimination among most popular cosmological models.
    The present-day constraints resulting from independent analyses
  by Maoli et al (\citenum{mao01}),  Rhodes et al (\citenum{rhodes01}),
  van Waerbeke et al
 (\citenum{vwal01,vw02}), Hoekstra et al
 (\citenum{hoek01a,hoekstra02})
  and R\'efr\'egier et al (\citenum{ref02})
   can be summarized by the following
  conservative  boundaries (90\% confidence level):
\begin{equation}
0.05 \le \Omega_m \le 0.8 \ \ \ \ {\rm and} \ \ \ \ 0.5 \le \sigma_8
\le 1.2 \ ,
\end{equation}
 and, in the case of a flat-universe with $\Omega_m=0.3$,  they
  lead to
  $\sigma_8 \approx 0.9$.
\\
\begin{figure}[t]
\begin{center}
\includegraphics[width=10cm]{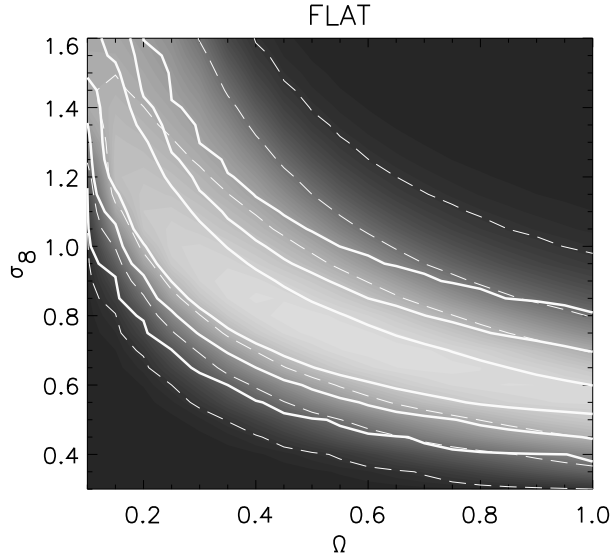}
\end{center}
\caption[]{Constraints on
  $\Omega$ and $\sigma_8$ for the flat cosmologies.  The confidence
  levels are $[68,95,99.9]$ from the brightest to the darkest area.
  The gray area and
  the dashed contours correspond to the computations with a full
  marginalisation over the default prior $\Gamma\in [0.05,0.7]$ and
  $z_s\in [0.24,0.64]$. The thick solid line contours are obtained from
 the prior $\Gamma\in [0.1,0.4]$ and $z_s\in [0.39,0.54]$ (which is a
  mean redshift $\bar z_s\in [0.8,1.1]$). From
  van Waerbeke et al. (2002) (\citenum{vw02}).
  }
\label{omegasig}
\end{figure}
\begin{figure}[t]
\begin{center}
\includegraphics[width=10cm]{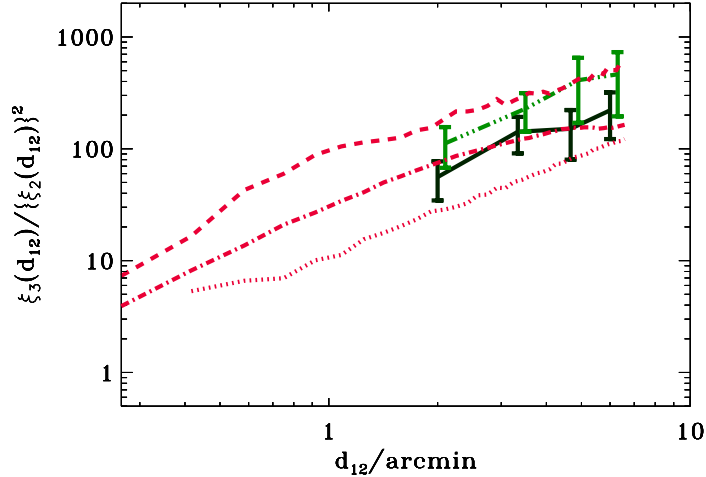}
\end{center}
\caption[]{Results for the VIRMOS-DESCART
 survey of the reduced three
point correlation function (\citenum{bern02b}). The solid line
 with error bars shows the raw results, when both the $E$ and $B$
contributions
to the two-point correlation functions are included.  The  dot-dashed
 line with error bars corresponds to measurements where the contribution
of the $B$
mode has been subtracted out from the two-point correlation function.
 These
measurements are compared to results obtained in $\tau $CDM, OCDM and
$\Lambda$CDM simulations (dashed, dotted and dot-dashed lines
respectively).  }
\label{xi3}
\end{figure}
\subsection{Toward breaking the
$\Omega_{{\rm m}}-\sigma_8$ degeneracy}
The measurement of non-Gaussian features needs informations
  on higher order statistics than variance. Eq.(\ref{eqs3}) 
  shows that  the  skewness of $\kappa$ seems the best suited 
  quantity for this purpose. However, in contrast with the variance, 
  the skewness demands first to reconstruct the mass map of the 
  field, which  complicates the process.  In fact, its measurements
   suffers from a number a practical difficulties which are not yet
    fixed.  In particular, the masking process generates a Swiss-Cheese 
  pattern over the field which significantly degrades the quality 
  of  the mass reconstruction and put strong doubts on the 
  reliability of its skewness and its cosmological interpretation.   
     Recently,
Bernardeau, van Waerbeke \& Mellier (\citenum{bern02a})  have
proposed an alternative method using some specific patterns in the shear
three-point function.
 Despite the complicated shape of the
three-point correlation pattern, they uncovered
 it can be used for the measurement
of non-Gaussian
 features. Their  detection strategy based on their method has
been tested on ray tracing simulations and
  turns out  to be robust, usable in patchy catalogs, and quite
insensitive to the topology of the survey.
\\
Bernardeau et al (\citenum{bern02b}) used the analysis
  of the  3-point correlations function on the VIRMOS-DESCART
data.  Their results on  Figure \ref{xi3} show a
2.4$\sigma$ signal over four independent angular bins, or
equivalently, a 4.9-$\sigma$ confidence level detection with respect
to measurements errors on scale of about $2$ to $4$ arc-minutes.
The amplitude and the shape of the signal are consistent
with theoretical expectations obtained from ray-tracing simulations.
This result supports
the idea that the measure corresponds to a  cosmological signal due to
the gravitational instability dynamics. Moreover,
  its properties could be used
to put constraints on the cosmological parameters, in particular on
the density parameter of the Universe. Although the
   errors
 are  still large to permit secure conclusions, one clearly see that
   the amplitude and the shape of the 3-point correlations function
   match the most likely cosmological models.  Remarkably, the
  $\Lambda$CDM scenario perfectly fit the data points.
\\
The Bernardeau et al (\citenum{bern02b}) result is the
first detection of non-Gaussian features in a cosmic shear survey and
  it opens the route to break the $\Omega_{{\rm m}}-\sigma_8$ degeneracy.
 Furthermore, this method is weakly dependent on other parameters,
  like the cosmological constant or the properties of the power
spectrum.
 However, there are still some caveats which may be considered
seriously.
 One difficulty is the source  clustering which could
   significantly perturb high-order statistics
 (Hamana et al 2000 (\citenum{hamana00})).
 If so, multi-lens plane cosmic shear analysis will be necessary
  which implies a good knowledge of the redshift distribution.
   For very deep cosmic shear surveys which 
  probe galaxies fainter than the limiting spectroscopic capabilities of 
  10-meter class telescopes, the lack of reliable 
  photometric redshift calibration  be  could be
  a serious concern.
\section{Outlook}
\begin{figure}[t]
\begin{center}
\includegraphics[width=8.5cm]{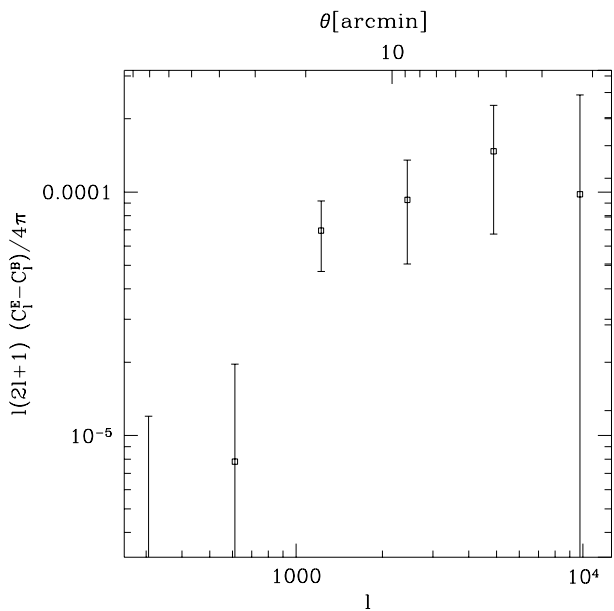}
\end{center}
\caption[]{Cosmological results from cosmic shear surveys:
 The angular power spectrum of the convergence field
  from the VIRMOS-DESCART  survey is plotted (From Pen et al 2001).
  These are the first $C(l)$ of dark matter ever measured in cosmology.}
\label{cl}
\end{figure}

 Because on going surveys increase
  both in solid angle and in number of galaxies, they
  will quickly  improve the accuracy of cosmic shear measurements, at
 a level where $\Omega_m$ and $\sigma_8$ will be known with  a
  10\% accuracy.   Since it is based on gravitational deflection
  by intervening matter spread over cosmological scales, the shape of
   the distortion field also probes directly the shape of the
  projected dark matter power spectrum. Pen et al (\citenum{pen01})
 explored its properties by measuring for the first
  time the $C(l)$ of the dark matter (see Figure
 \ref{cl}).  Although the reliability of these 
   $C(l)$ is still unknown, this work shows 
   this is feasible with present data.  The capabilities of
 cosmic shear surveys to probe dark matter can be also used to explore the
   relations between light and mass at different angular scale,  
  different redshifts  and in different environment. Recently 
   Hoekstra et al (\citenum{hoeskbias2}) joined together the 
 RCS and the VIRMOS-DESCART surveys and computed both the biasing and 
  the galaxy-mass cross-correlation. They used 
   the aperture mass and aperture number density statistics, 
  as it was originally proposed by Schneider (\citenum{schb98})
   and van Waerbeke (\citenum{ludobias98}).  
   They found that
  the biasing is linear on scale beyond 3 $h^{-1}$, but turns out to be 
more complex on lower scale, indicating a significant non-linear and/or 
  stochastic biasing. These results are indeed difficult to interpret 
  in a cosmological context without high resolution numerical simulations 
  and a detailed scenario of galaxy formation in hands. But they 
  show the remarkable potential of cosmic shear survey for challenging 
   theory of galaxy formation against observations. Moreover, 
  it is worth noticing that this work was made possible only because 
  the two surveys merged their data in order to cover a sufficiently large
  field of view.  The investigation of mass and light relations 
   is certainly 
  one of the most promising goals of  wide cosmic shear surveys in the future.
   
\begin{figure}[t]
\begin{center}
\includegraphics[width=9cm]{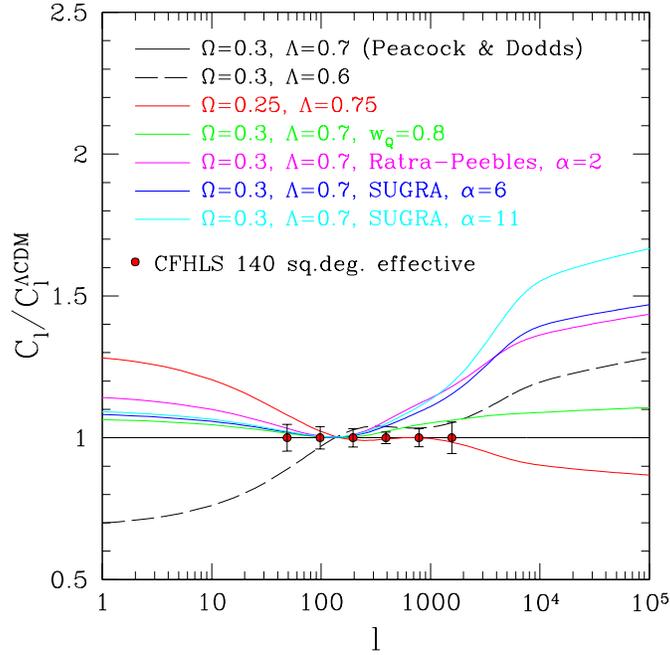}
\end{center}
\caption[]{The future of from cosmic shear surveys:
  Theoretical expectations of the CFHT Legacy Survey. The dots with
error
bars are the expected measurements of cosmic shear data from the 208
deg$^2$ of the CFHT Legacy Survey. The lines shows various models
  discussed by Benabed \& Bernardeau (\citenum{benab01}).}
\label{future1}
\end{figure}

However, we expect much more within the next decade.  Several  
  new wide field surveys 
  are in progress at ESO/La Silla (Erben et al 2002 in preparation), 
  at CTIO (Jarvis et al), at NOAO (Januzzi et al) and 
   at SUBARU (Miyazaki 2002 (\citenum{subashear})). Surveys
  covering
   hundreds of degrees, with multi-bands data in order
to get redshift of sources and possibly detailed information of
 their clustering properties, are scheduled.  The CFHT Legacy
Survey\footnote{http://www.cfht.hawaii.edu/Science/CFHLS/} will cover
  200 deg$^2$ and is one of those
next-generation cosmic shear survey. Figures 1 and \ref{future1} shows
 their expected  
  potential for cosmology. On figure 1 we simulated the expected signal
to noise of the shear variance as function of angular scale for
  a $\Lambda$CDM cosmology.  The error bars are considerably reduced as
compared
 to present-day survey.  On Fig. \ref{future1}, we compare the expected
signal to noise
 ratio of the CFHT Legacy Survey with the expected amplitude of the
angular
power spectrum for several theoretical quintessence fields models. It
shows
that even with 200 deg$^2$ which include multi-color informations in
order to
get redshift of sources, one can already obtain interesting constraints
on
  cosmology beyond standard models.\\
The use of cosmic shear data can be much more efficient if they are used
together
  with other surveys, like CMB (Boomerang, MAP, Planck), SNIa surveys,
or even
galaxy surveys (2dF, SDSS).  For example, SDSS will soon provide
  the 100, 000 quasars with redshifts.  M\'enard \& Bartelmann
 (\citenum{menbart02})
    have
recently explored the
   interest of this survey in order to cross-correlate the foreground
 galaxy distribution with the quasar population.  The expected
magnification
  bias generated by dark matter associated with foreground
  structures as mapped by galaxies  depends on $\Omega_m$
  and the biasing $\sigma_8$.
    In principle magnification bias in
the
SDSS quasar sample can provide similar constrains as cosmic
shear. Both cosmic shear and cosmic magnification 
  wide field surveys should soon produce impressive constraints on cosmological   models.
\section*{Aknowledgements}
 We thank M. Bartelmann, R. Carlberg, T. Erben, B. Fort, H. Hoekstra, B. 
 Jannuzi, B.,  M\'enard,
 O. Dor\'e,  U.-L. Pen, S. Prunet and P. Schneider  for useful
   discussions.
   This work was supported by the TMR Network
``Gravitational
 Lensing: New Constraints on
Cosmology and the Distribution of Dark Matter'' of the EC under contract
No. ERBFMRX-CT97-0172. 


\end{document}